\documentclass[prl,twocolumn,superscriptaddress,epsfig]{revtex4}
\usepackage{graphicx}
\usepackage{dcolumn}
\usepackage{bm}
\begin{document}
\title {Thermoelectric Properties of Electrostatically Tunable Antidot Lattices}
\author{Srijit~Goswami}
\email{sg483@cam.ac.uk}
\affiliation{Cavendish Laboratory, University of Cambridge, J.J. Thomson Avenue, Cambridge CB3 0HE, United Kingdom.}
\author{Christoph~Siegert}
\affiliation{Cavendish Laboratory, University of Cambridge, J.J. Thomson Avenue, Cambridge CB3 0HE, United Kingdom.}
\author{Saquib~Shamim}
\affiliation{Department of Physics, Indian Institute of Science, Bangalore 560 012, India.}
\author{Michael~Pepper}
\thanks{Present address: Department of Electronic and Electrical Engineering, University College, London.}
\affiliation{Cavendish Laboratory, University of Cambridge, J.J. Thomson Avenue, Cambridge CB3 0HE, United Kingdom.}
\author{Ian~Farrer}
\affiliation{Cavendish Laboratory, University of Cambridge, J.J. Thomson Avenue, Cambridge CB3 0HE, United Kingdom.}
\author{David~A.~Ritchie}
\affiliation{Cavendish Laboratory, University of Cambridge, J.J. Thomson Avenue, Cambridge CB3 0HE, United Kingdom.}
\author{Arindam~Ghosh}
\affiliation{Department of Physics, Indian Institute of Science, Bangalore 560 012, India.}
\begin{abstract}
We report on the fabrication and characterization of a device which allows the formation of an antidot lattice (ADL) using only electrostatic gating. The antidot potential and Fermi energy of the system can be tuned independently. Well defined commensurability features in magnetoresistance as well as magnetothermopower are obsereved. We show that the thermopower can be used to efficiently map out the potential landscape of the ADL.
\end{abstract}


\maketitle

A two dimensional electron system (2DES) with a regular array of scatterers is referred to as an antidot lattice (ADL). Over the past two decades, such systems have been studied in great detail revealing a variety of intriguing physical phenomena, ranging from classical pinned orbits in magnetic fields (commensurability)~\cite{PhysRevLett.66.2790} to Ahoronov-Bohm type quantum interference effects~\cite{PhysRevB.49.8510}. Recently, there has been a renewed interest in ADLs, as it has been shown that they can alter the band structure of a system, resulting in modified optical~\cite{PhysRevB.77.245431} and electrical~\cite{NJP.11.095020} properties. ADLs are usually fabricated by the physical removal of material (etching) in the desired region, resulting in a fixed potential profile in the underlying 2DES~\cite{PhysRevLett.66.2790}. Less destructive techniques such as local oxidation~\cite{PhysRevB.71.035343} or electrostatic gating~\cite{PhysRevB.44.3447} have also been employed in the fabrication of ADLs. However, none of the existing techniques allow for an independent variation of the antidot potential $(U_{AD})$ and the Fermi energy $(E_{F})$ of the electron sea between the antidots. The fabrication of such devices with independent tunability of $U_{AD}$ and $E_{F}$, would allow for a systematic study of the interplay between these two relevant energy scales. In view of recent proposals for scalable spin-based quantum information processing devices in ADLs~\cite{PhysRevLett.100.136804,NL.5.2515}, this tunability is highly desirable. Furthermore, recent studies in 2D mesoscopic devices in  GaAs/AlGaAs heterostructures have revealed the possibility of the existence of spin correlated systems formed due to the existence of intrinsic potential modulations within the 2DES~\cite{NP.3.315}. Such tunable devices provide the possibility of realizing well-ordered artificial spin-lattices with tunable magnetism in a semiconductor environment.
\begin{figure}[tbh]
\includegraphics[width=1\linewidth]{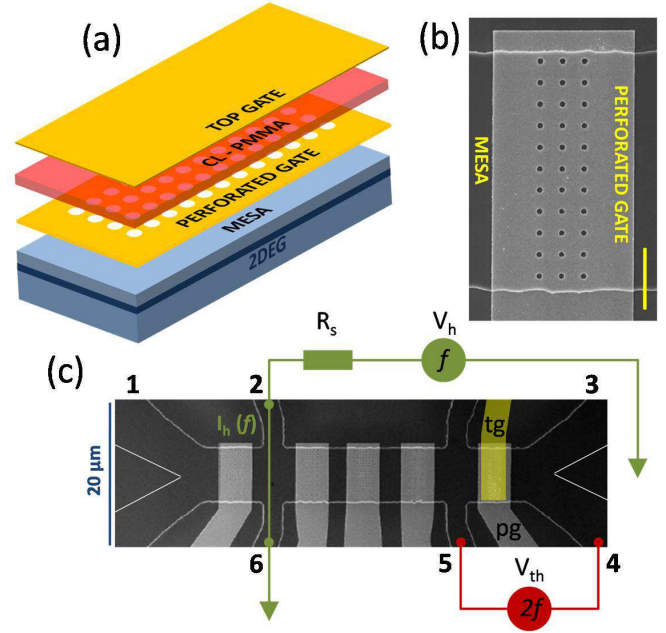}
\caption{(a) Exploded view of the device structure.  (b) Scanning electron micrograph of a device similar to the ones investigated in this work. The scale bar is $2$~$\mu$m. (c) Schematic of the measurement set-up used to study thermopower (superimposed on an SEM image showing a number of devices on a single chip before the second metallization).}
\end{figure}
In this Letter, we describe the properties of ADLs which are induced solely by electrostatic gating. We perform magnetoresistance (MR) and magnetothermopower (MTP) studies to show that $U_{AD}$ and $E_{F}$ can be controlled independently. We also show that thermopower (TP) is an extremely sensitive tool to study the potential landscape of the ADL system. Silicon $\delta$-doped GaAs/AlGaAs heterostructures with a $30$~nm spacer layer with an as-grown number density of $4\times 10^{15}$~m$^{-2}$, and low temperature mobility of $230$~m$^{2}$/Vs (corresponding to an elastic mean free path $\sim 20$~$\mu$m ) were used in our experiments.

Figure~1(a) shows a schematic of the device structure. Following the definition of a mesa by wet etching, standard electron beam lithography and lift-off techniques were used to obtain a metallic (Ti/Au) perforated gate (PG). The antidot diameter~$(d)$ and spacing~$(a)$ are fixed by the geometry of the perforations. A $150$~nm thick layer of cross-linked polymethyl methacrylate serves as a dielectric between the PG and a Ti/Au top gate (TG). Application of suitable negative voltages $(V_{tg})$ on the TG result in the formation of a perfectly periodic ADL in the 2DES. The sheet density, $n_{s}$ (or $E_{F}$), of the itinerant electrons between the scatterers can be continuously varied by applying voltages $(V_{pg})$ to the the PG. Figure~1(b) shows a scanning electron micrograph of one such PG (similar to the ones used in this experiment) prior to deposition of the TG. We were able to reliably fabricate devices with $d$($a$) ranging from $80$~nm to $300$~nm ($160$~nm to $700$~nm).

Measurements were performed in a dilution refrigerator with a base temperature of $30$~mK. For resistance measurements a current ($I=1$~$\mu$A, frequency $(f)$~$=11.3$~Hz) was passed between contacts 1 and 3 [Figure~1(c)], and the resultant voltage drop across the device (dev1:~$d=300$~nm, $a=700$~nm) was measured between contacts 4 and 5 using a lock-in amplifier yielding the four-probe resistance $(R)$ of the device. Under the influence of a perpendicular magnetic field $(B)$, electrons move in cyclotron orbits and the cyclotron radius $(r_{c})$ is given by $r_{c}=\frac{\hbar\sqrt{2\pi n_{s}}}{eB}$, where $e$ is the electron charge. At particular values of $B$, electrons move in pinned orbits around a specific number of antidots resulting in peaks in the resistance whenever $r_{c}$ is commensurate with the lattice~\cite{PhysRevLett.66.2790}. Figure 2(b) shows commensurability features for two devices (dev1: $d=300$~nm, $a=700$~nm; dev2: $d=200$~nm, $a=400$~nm) at 4.2 K. $V_{tg}$ was fixed at ($-35$~V~/$-50$~V) for (dev1/dev2) with $V_{pg} = 0$~V resulting in the formation of a strong periodic antidot potential in the 2DES~\cite{note1}. Clear commensurabilty peaks are observed in both devices corresponding to pinned orbits around $1$ (fundamental), $4$ and $9$ antidots. The higher order peaks ($4$ and $9$) shift to lower $B$ for dev2. This shift has been reported previously for ADLs with a high aspect ratio (i.e., $d/a$)~\cite{PhysRevLett.66.2790} and is attributed to chaotic electron dynamics~\cite{PhysRevLett.68.1367}. In contrast, we see no such shift for dev1, which has a smaller aspect ratio. We note that dev1 consists of $33$ antidots (in a $3\times 11$~array), which is much lower than typical studies~\cite{PhysRevLett.66.2790,PhysRevB.71.035343}. Despite this, we are able to clearly resolve even higher order commensurability features. This may be due to the fact that the non-destructive (gating) technique used to form the ADL does not severely alter the local electron mobility. For the rest of this work we focus on a detailed study of dev1.

Figure~2(c) shows the variation of low temperature ($30$~mK) MR as a function of $V_{tg}$. As $V_{tg}$ is reduced (i.e., $U_{AD}$ is increased) from $-20$~V to $-35$~V in steps of $1$~V (with $V_{pg}$ held at $0$~V), we see a clear emergence of commensurability features (for $V_{tg}\lesssim -25$~V), which increase in strength as $V_{tg}$ is reduced. This is illustrated schematically in Figure~2(a1 and a2). $n_{s}$ can be determined from the position of the fundamental commensurability peak ($B=0.297$~T), and yields a value of $n_{s}=3.96\times 10^{15}$~m$^{-2}$. Alternatively, $n_{s}$ may be extracted directly from the low-$B$ Shubnikov de-Haas (SdH) oscillations resulting in $n_{s}=3.91\times 10^{15}$~m$^{-2}$, in excellent agreement with that extracted from the commensurability condition. Using this value of $n_{s}$ we compute $r_{c}$ corresponding to the $B=0.139$~T and $B=0.110$~T and draw the resultant cyclotron orbits [inset of Figure~2(c)], which confirm that they correspond to orbits around $4$ and $9$ antidots. We emphasize that the position of the peak is independent of $V_{tg}$ in this range, thus demonstrating that $U_{AD}$ may be continuously varied without any significant effect on $E_{F}$ of the electron sea. Figure~2(d) shows that at a fixed $V_{tg}$ ($=-35$~V), $E_{F}$ can be systematically tuned without affecting $U_{AD}$. As $V_{pg}$ is varied from $0$~V to $-0.36$~V [illustrated in Figure~2(a3)], the position of the peak changes continuously from $B=0.297$~T to $B=0.282$~T, without a noticeable change in MR profile. This corresponds to a change in $n_{s}$ from  $3.96\times 10^{15}$~m$^{-2}$ to $3.57\times 10^{15}~$m$^{-2}$ (i.e., $\Delta E_{F}=1.4$~meV). Thus, it is clear that $U_{AD}$ and $E_{F}$ can indeed be controlled independently in our ADL devices.

\begin{figure}[tb]
\includegraphics[width=1\linewidth]{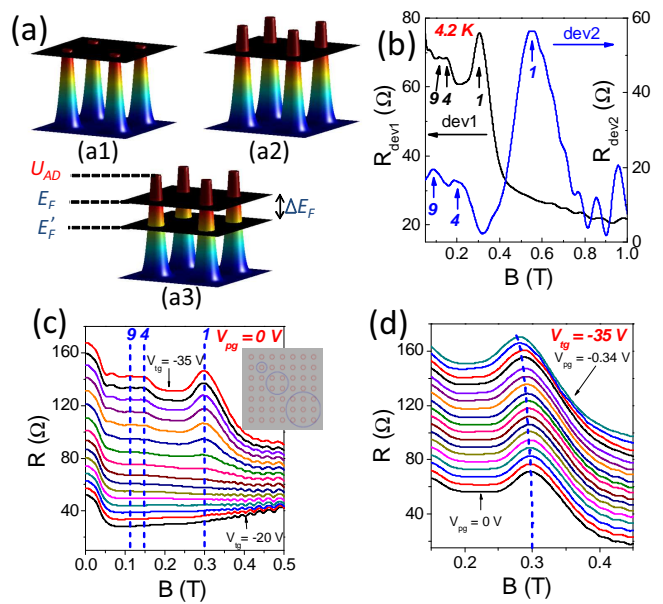}
\caption{(a) Schematic of the relevant energy scales in the ADL system: (a1) A situation when $U_{AD}$ just pushes through $E_{F}$, (a2) a relatively high $U_{AD}$ resulting in strong scattering centres, (a3) variation of $E_{F}$ with a fixed $U_{AD}$. (b) Commensurability in two devices, dev1 ($d=300$~nm, $a=700$~nm), and dev2 ($d=200$~nm, $a=400$~nm) at $4.2$~K. Arrows mark the position of commensurability peaks corresponding to cyclotron orbits around $1$, $4$, and $9$ antidots. (c) MR as $V_{tg}$ is varied from $-35$~V to $-20$~V in steps of $1$~V. Curves have been offset by $5$~$\Omega$ for clarity. (d) Variation in the fundamental commensurability peak with $V_{pg}$, showing a clear modulation of $E_{F}$ (offset$=5$~$\Omega$). Dashed (blue) line serves as a guide to the eye.}
\end{figure}

We now turn our attention to the thermopower (TP) of the ADL. TP is defined as $S=\lim_{\Delta T \to 0} \frac{V_{th}}{\Delta T}$, where $\Delta T$ is the temperature difference across the system, and $V_{th}$ is the resultant thermovoltage. TP often yields vital information about energy flow through the system that cannot be gleaned from standard resistance and conductance measurements~\cite{PhysRevB.62.R16275,PhysRevLett.95.176602}. Various studies in low-dimensional systems have shown that TP is highly sensitive to the local density of states~\cite{PhysRevLett.103.026602,PhysRevB.62.R16275,PhysRevLett.95.176602}. However there have been extremely limited studies of TP in ADLs~\cite{JETP.79.166,JETP.81.462}. We show that TP is an excellent tool to systematically study ADLs. A schematic of the set-up used for TP measurements is shown in Figure~1(c). A local heating technique is used where a heating current ($I_{h}=10$~$\mu$A, $f=11.3$~Hz) flows through one arm of the mesa resulting in a specific $\Delta T$ across the device. $V_{th}$ is measured (between contacts 4 and 5) using a lock-in amplifier at $2f$. Note that due to the lack of local temperature sensors, the current device geometry does not allow for an accurate measurement of $\Delta T$ and hence $S$. However, the underlying physics can be explored equally well through a measurement of $V_{th}$.

\begin{figure}[tb]
\includegraphics[width=1\linewidth]{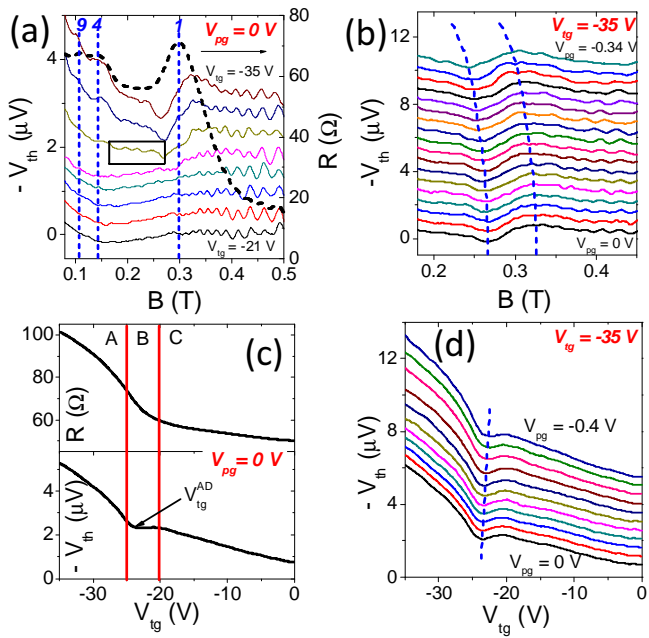}
\caption{(a) Variation of MTP as $V_{tg}$ is changed from $-35$~V to $-21$~V ($V_{pg}=0$~V). Dashed (black) line shows the MR at $V_{tg}=-35$~V for comparison. (offset$=0.6$~$\mu$V) (b) A systematic shift in position of the fundamental commensurability feature in MTP. Dashed (blue) lines are guides to the eye. (c) (upper/lower) panel shows ($R$/$-V_{th}$) \emph{vs.} $V_{tg}$. A, B, and C show three distinct transport regimes in the device. (d) Shift in position of $V_{tg}^{AD}$ with $V_{pg}$ (offset$=0.6$~$\mu$V).}
\end{figure}

Figure 3(a) shows the variation of MTP with $V_{tg}$. We find that the signature of commensurability in TP is a sharp change in its magnitude. This can be seen clearly when compared with the peak observed in MR [dashed (black) line in Figure 3(a)]. This is a direct consequence of the Mott formula, appropriately modified for an array of scatterers~\cite{PhysRevB.66.201303}. The qualitative behavior of TP in our device agrees well with that expected from ref. 17 (not shown). As expected, the strength of the feature diminishes when $V_{tg}$ changes from $-35$~V to $-21$~V. Figure 3(b) clearly shows a shift in its position when $V_{tg}$ is held constant and $V_{pg}$ is changed from $0$~V to $-0.34$~V. In addition to the $(1/B)$-periodic SdH oscillations, we observe highly reproducible oscillations in TP at low $B$ ($<0.3$~T), which are periodic in $B$ [boxed region in Figure~3(a)]. Similar oscillations in MR have been reported earlier and have been associated with quantum interference phenomena in ADLs~\cite{PhysRevB.51.4649} or anti-dot charging effects in single antidots~\cite{PhysRevLett.83.160}. Budantsev \emph{et al.}~\cite{JETP.79.166} have observed seemingly similar oscillations in the \emph{transverse} TP, and attribute them to interference due to electron orbits \emph{between} four neighboring antidots. However, the observed TP oscillations in our case are associated with electron motion \emph{around} a single antidot. A detailed analysis of these oscillations will be the subject of another work. We note that these B-periodic oscillations are clearly absent in MR, reinforcing the fact that TP is a more sensitive tool to probe the electronic structure of ADLs.

We now show that the sensitivity of TP can be exploited to directly investigate the interplay between $U_{AD}$ and $E_{F}$. Transport through the ADL can be divided into three regimes (A,B and C) [Figure~3(c)]. As $V_{tg}$ is reduced from $0$~V to $\sim -20$~V (region C), the local number density (in the regions determined by the perforations) gradually changes, resulting in an increase in $R$ (upper pannel) and $-V_{th}$ (lower pannel). Region B shows a rapid increase in $R$ and a sharp dip in $-V_{th}$. This indicates the onset of strong scattering centers in the 2DES (i.e., the formation of the ADL). In other words, in this region $U_{AD}$ just \emph{pierces} the Fermi surface [illustrated in Figure~2(a1)], and electrons see an array of scatterers. This picture is corroborated by the fact that commensurability features in MR (and MTP) only become visible when $V_{tg}<-20$~V. For $V_{tg}<-25$~V (region A), sidewise depletion becomes important, resulting in a further increase in TP due to an increase in $U_{AD}$ [illustrated in Figure~2(a2)]. Thus, region B corresponds to a crossover from a situation when the ADL is not defined to one where $U_{AD}>E_{F}$. The position of the minimum in $-V_{th}$ ($V_{tg}^{AD}$) may thus be used to identify region B. In this picture, one would expect that $V_{tg}^{AD}$ should be highly sensitive to $E_{F}$. Figure 3(d) shows the variation of $-V_{th}$ with $V_{tg}$ for various $V_{pg}$. As $E_{F}$ decreases $V_{tg}^{AD}$ clearly moves to less negative values, indicating that the ADL is induced at lower values of $|V_{tg}|$. The non-monotonic behaviour of TP in region B is a purely electrostatic phenomenon, found to be practically independent of temperature. It may be related to inhomogeneties in $U_{AD}$ resulting from fluctuations in the conduction band due to rapid variations in the impurity potential~\cite{NP.3.315}. However, a more detailed study is required to confirm its origin. Despite this, we can clearly see that the TP is highly sensitive to the relative magnitudes of $E_{F}$ and $U_{AD}$, and can be used to accurately determine the region in which the ADL is formed.

In conclusion, we have demonstrated an ADL device where the antidot potential and Fermi energy of the system can be controlled independently through electrostatic gating. We have characterized the device using resistance and TP measurements and show that TP is a particularly sensitive tool to study the potential landscape in ADLs.

This work was supported by the Department of Science and Technology (Govt. of India), EPSRC (U.K.), and UK-India Educational Research Initiative (UKIERI). S.G. would like to thank the Gates Cambridge Trust for financial support.


\begin{thebibliography}{1}
\expandafter\ifx\csname natexlab\endcsname\relax\def\natexlab#1{#1}\fi
\expandafter\ifx\csname bibnamefont\endcsname\relax
  \def\bibnamefont#1{#1}\fi
\expandafter\ifx\csname bibfnamefont\endcsname\relax
  \def\bibfnamefont#1{#1}\fi
\expandafter\ifx\csname citenamefont\endcsname\relax
  \def\citenamefont#1{#1}\fi
\expandafter\ifx\csname url\endcsname\relax
  \def\url#1{\texttt{#1}}\fi
\expandafter\ifx\csname urlprefix\endcsname\relax\def\urlprefix{URL }\fi
\providecommand{\bibinfo}[2]{#2}
\providecommand{\eprint}[2][]{\url{#2}}

\bibitem[{\citenamefont{Weiss et~al.}(1991)\citenamefont{Weiss, Roukes,
  Menschig, Grambow, von Klitzing, and Weimann}}]{PhysRevLett.66.2790}
\bibinfo{author}{\bibfnamefont{D.}~\bibnamefont{Weiss}},
  \bibinfo{author}{\bibfnamefont{M.~L.} \bibnamefont{Roukes}},
  \bibinfo{author}{\bibfnamefont{A.}~\bibnamefont{Menschig}},
  \bibinfo{author}{\bibfnamefont{P.}~\bibnamefont{Grambow}},
  \bibinfo{author}{\bibfnamefont{K.}~\bibnamefont{von Klitzing}},
  \bibnamefont{and} \bibinfo{author}{\bibfnamefont{G.}~\bibnamefont{Weimann}},
  \bibinfo{journal}{Phys. Rev. Lett.} \textbf{\bibinfo{volume}{66}},
  \bibinfo{pages}{2790} (\bibinfo{year}{1991}).

\bibitem[{\citenamefont{Schuster et~al.}(1994)\citenamefont{Schuster, Ensslin,
  Wharam, K\"uhn, Kotthaus, B\"ohm, Klein, Tr\"ankle, and
  Weimann}}]{PhysRevB.49.8510}
\bibinfo{author}{\bibfnamefont{R.}~\bibnamefont{Schuster}},
  \bibinfo{author}{\bibfnamefont{K.}~\bibnamefont{Ensslin}},
  \bibinfo{author}{\bibfnamefont{D.}~\bibnamefont{Wharam}},
  \bibinfo{author}{\bibfnamefont{S.}~\bibnamefont{K\"uhn}},
  \bibinfo{author}{\bibfnamefont{J.~P.} \bibnamefont{Kotthaus}},
  \bibinfo{author}{\bibfnamefont{G.}~\bibnamefont{B\"ohm}},
  \bibinfo{author}{\bibfnamefont{W.}~\bibnamefont{Klein}},
  \bibinfo{author}{\bibfnamefont{G.}~\bibnamefont{Tr\"ankle}},
  \bibnamefont{and} \bibinfo{author}{\bibfnamefont{G.}~\bibnamefont{Weimann}},
  \bibinfo{journal}{Phys. Rev. B} \textbf{\bibinfo{volume}{49}},
  \bibinfo{pages}{8510} (\bibinfo{year}{1994}).

\bibitem[{\citenamefont{Pedersen
  et~al.}(2008{\natexlab{a}})\citenamefont{Pedersen, Flindt, Pedersen, Jauho,
  Mortensen, and Pedersen}}]{PhysRevB.77.245431}
\bibinfo{author}{\bibfnamefont{T.~G.} \bibnamefont{Pedersen}},
  \bibinfo{author}{\bibfnamefont{C.}~\bibnamefont{Flindt}},
  \bibinfo{author}{\bibfnamefont{J.}~\bibnamefont{Pedersen}},
  \bibinfo{author}{\bibfnamefont{A.-P.} \bibnamefont{Jauho}},
  \bibinfo{author}{\bibfnamefont{N.~A.} \bibnamefont{Mortensen}},
  \bibnamefont{and} \bibinfo{author}{\bibfnamefont{K.}~\bibnamefont{Pedersen}},
  \bibinfo{journal}{Phys. Rev. B} \textbf{\bibinfo{volume}{77}},
  \bibinfo{pages}{245431} (\bibinfo{year}{2008}{\natexlab{a}}).

\bibitem[{\citenamefont{F\"{u}rst et~al.}(2009)\citenamefont{F\"{u}rst,
  Pedersen, Flindt, Mortensen, Brandbyge, Pedersen, and Jauho}}]{NJP.11.095020}
\bibinfo{author}{\bibfnamefont{J.~A.} \bibnamefont{F\"{u}rst}},
  \bibinfo{author}{\bibfnamefont{J.~G.} \bibnamefont{Pedersen}},
  \bibinfo{author}{\bibfnamefont{C.}~\bibnamefont{Flindt}},
  \bibinfo{author}{\bibfnamefont{N.~A.} \bibnamefont{Mortensen}},
  \bibinfo{author}{\bibfnamefont{M.}~\bibnamefont{Brandbyge}},
  \bibinfo{author}{\bibfnamefont{T.~G.} \bibnamefont{Pedersen}},
  \bibnamefont{and} \bibinfo{author}{\bibfnamefont{A.-P.} \bibnamefont{Jauho}},
  \bibinfo{journal}{New Journal of Physics} \textbf{\bibinfo{volume}{11}},
  \bibinfo{pages}{095020} (\bibinfo{year}{2009}).

\bibitem[{\citenamefont{Dorn et~al.}(2005)\citenamefont{Dorn, Bieri, Ihn,
  Ensslin, Driscoll, and Gossard}}]{PhysRevB.71.035343}
\bibinfo{author}{\bibfnamefont{A.}~\bibnamefont{Dorn}},
  \bibinfo{author}{\bibfnamefont{E.}~\bibnamefont{Bieri}},
  \bibinfo{author}{\bibfnamefont{T.}~\bibnamefont{Ihn}},
  \bibinfo{author}{\bibfnamefont{K.}~\bibnamefont{Ensslin}},
  \bibinfo{author}{\bibfnamefont{D.~D.} \bibnamefont{Driscoll}},
  \bibnamefont{and} \bibinfo{author}{\bibfnamefont{A.~C.}
  \bibnamefont{Gossard}}, \bibinfo{journal}{Phys. Rev. B}
  \textbf{\bibinfo{volume}{71}}, \bibinfo{pages}{035343}
  (\bibinfo{year}{2005}).

\bibitem[{\citenamefont{Lorke et~al.}(1991)\citenamefont{Lorke, Kotthaus, and
  Ploog}}]{PhysRevB.44.3447}
\bibinfo{author}{\bibfnamefont{A.}~\bibnamefont{Lorke}},
  \bibinfo{author}{\bibfnamefont{J.~P.} \bibnamefont{Kotthaus}},
  \bibnamefont{and} \bibinfo{author}{\bibfnamefont{K.}~\bibnamefont{Ploog}},
  \bibinfo{journal}{Phys. Rev. B} \textbf{\bibinfo{volume}{44}},
  \bibinfo{pages}{3447} (\bibinfo{year}{1991}).

\bibitem[{\citenamefont{Pedersen
  et~al.}(2008{\natexlab{b}})\citenamefont{Pedersen, Flindt, Pedersen,
  Mortensen, Jauho, and Pedersen}}]{PhysRevLett.100.136804}
\bibinfo{author}{\bibfnamefont{T.~G.} \bibnamefont{Pedersen}},
  \bibinfo{author}{\bibfnamefont{C.}~\bibnamefont{Flindt}},
  \bibinfo{author}{\bibfnamefont{J.}~\bibnamefont{Pedersen}},
  \bibinfo{author}{\bibfnamefont{N.~A.} \bibnamefont{Mortensen}},
  \bibinfo{author}{\bibfnamefont{A.-P.} \bibnamefont{Jauho}}, \bibnamefont{and}
  \bibinfo{author}{\bibfnamefont{K.}~\bibnamefont{Pedersen}},
  \bibinfo{journal}{Phys. Rev. Lett.} \textbf{\bibinfo{volume}{100}},
  \bibinfo{pages}{136804} (\bibinfo{year}{2008}{\natexlab{b}}).

\bibitem[{\citenamefont{Flindt et~al.}(2005)\citenamefont{Flindt, Mortensen,
  and Jauho}}]{NL.5.2515}
\bibinfo{author}{\bibfnamefont{C.}~\bibnamefont{Flindt}},
  \bibinfo{author}{\bibfnamefont{N.~A.} \bibnamefont{Mortensen}},
  \bibnamefont{and} \bibinfo{author}{\bibfnamefont{A.-P.} \bibnamefont{Jauho}},
  \bibinfo{journal}{Nano Letters} \textbf{\bibinfo{volume}{5}},
  \bibinfo{pages}{2515} (\bibinfo{year}{2005}).

\bibitem[{\citenamefont{Siegert et~al.}(2007)\citenamefont{Siegert, Ghosh,
  Pepper, Farrer, and Ritchie}}]{NP.3.315}
\bibinfo{author}{\bibfnamefont{C.}~\bibnamefont{Siegert}},
  \bibinfo{author}{\bibfnamefont{A.}~\bibnamefont{Ghosh}},
  \bibinfo{author}{\bibfnamefont{M.}~\bibnamefont{Pepper}},
  \bibinfo{author}{\bibfnamefont{I.}~\bibnamefont{Farrer}}, \bibnamefont{and}
  \bibinfo{author}{\bibfnamefont{D.~A.} \bibnamefont{Ritchie}},
  \bibinfo{journal}{Nature Phys.} \textbf{\bibinfo{volume}{3}},
  \bibinfo{pages}{315} (\bibinfo{year}{2007}).

\bibitem[{not()}]{note1}
\bibinfo{note}{$R_{\textrm{dev2}}$ is obtained after subtracting a linear
  background to enhance the clarity of the commensurability features.}

\bibitem[{\citenamefont{Fleischmann et~al.}(1992)\citenamefont{Fleischmann,
  Geisel, and Ketzmerick}}]{PhysRevLett.68.1367}
\bibinfo{author}{\bibfnamefont{R.}~\bibnamefont{Fleischmann}},
  \bibinfo{author}{\bibfnamefont{T.}~\bibnamefont{Geisel}}, \bibnamefont{and}
  \bibinfo{author}{\bibfnamefont{R.}~\bibnamefont{Ketzmerick}},
  \bibinfo{journal}{Phys. Rev. Lett.} \textbf{\bibinfo{volume}{68}},
  \bibinfo{pages}{1367} (\bibinfo{year}{1992}).

\bibitem[{\citenamefont{Appleyard et~al.}(2000)\citenamefont{Appleyard,
  Nicholls, Pepper, Tribe, Simmons, and Ritchie}}]{PhysRevB.62.R16275}
\bibinfo{author}{\bibfnamefont{N.~J.} \bibnamefont{Appleyard}},
  \bibinfo{author}{\bibfnamefont{J.~T.} \bibnamefont{Nicholls}},
  \bibinfo{author}{\bibfnamefont{M.}~\bibnamefont{Pepper}},
  \bibinfo{author}{\bibfnamefont{W.~R.} \bibnamefont{Tribe}},
  \bibinfo{author}{\bibfnamefont{M.~Y.} \bibnamefont{Simmons}},
  \bibnamefont{and} \bibinfo{author}{\bibfnamefont{D.~A.}
  \bibnamefont{Ritchie}}, \bibinfo{journal}{Phys. Rev. B}
  \textbf{\bibinfo{volume}{62}}, \bibinfo{pages}{R16275}
  (\bibinfo{year}{2000}).

\bibitem[{\citenamefont{Scheibner et~al.}(2005)\citenamefont{Scheibner,
  Buhmann, Reuter, Kiselev, and Molenkamp}}]{PhysRevLett.95.176602}
\bibinfo{author}{\bibfnamefont{R.}~\bibnamefont{Scheibner}},
  \bibinfo{author}{\bibfnamefont{H.}~\bibnamefont{Buhmann}},
  \bibinfo{author}{\bibfnamefont{D.}~\bibnamefont{Reuter}},
  \bibinfo{author}{\bibfnamefont{M.~N.} \bibnamefont{Kiselev}},
  \bibnamefont{and} \bibinfo{author}{\bibfnamefont{L.~W.}
  \bibnamefont{Molenkamp}}, \bibinfo{journal}{Phys. Rev. Lett.}
  \textbf{\bibinfo{volume}{95}}, \bibinfo{pages}{176602}
  (\bibinfo{year}{2005}).

\bibitem[{\citenamefont{Goswami et~al.}(2009)\citenamefont{Goswami, Siegert,
  Baenninger, Pepper, Farrer, Ritchie, and Ghosh}}]{PhysRevLett.103.026602}
\bibinfo{author}{\bibfnamefont{S.}~\bibnamefont{Goswami}},
  \bibinfo{author}{\bibfnamefont{C.}~\bibnamefont{Siegert}},
  \bibinfo{author}{\bibfnamefont{M.}~\bibnamefont{Baenninger}},
  \bibinfo{author}{\bibfnamefont{M.}~\bibnamefont{Pepper}},
  \bibinfo{author}{\bibfnamefont{I.}~\bibnamefont{Farrer}},
  \bibinfo{author}{\bibfnamefont{D.~A.} \bibnamefont{Ritchie}},
  \bibnamefont{and} \bibinfo{author}{\bibfnamefont{A.}~\bibnamefont{Ghosh}},
  \bibinfo{journal}{Phys. Rev. Lett.} \textbf{\bibinfo{volume}{103}},
  \bibinfo{pages}{026602} (\bibinfo{year}{2009}).

\bibitem[{\citenamefont{Budantsev et~al.}(2004)\citenamefont{Budantsev, Lavrov,
  Pogosov, Plotnikov, Bakarov, Toropov, Maude, and Portal}}]{JETP.79.166}
\bibinfo{author}{\bibfnamefont{M.~V.} \bibnamefont{Budantsev}},
  \bibinfo{author}{\bibfnamefont{R.~A.} \bibnamefont{Lavrov}},
  \bibinfo{author}{\bibfnamefont{A.~G.} \bibnamefont{Pogosov}},
  \bibinfo{author}{\bibfnamefont{A.~E.} \bibnamefont{Plotnikov}},
  \bibinfo{author}{\bibfnamefont{A.~K.} \bibnamefont{Bakarov}},
  \bibinfo{author}{\bibfnamefont{A.~I.} \bibnamefont{Toropov}},
  \bibinfo{author}{\bibfnamefont{D.~K.} \bibnamefont{Maude}}, \bibnamefont{and}
  \bibinfo{author}{\bibfnamefont{J.~C.} \bibnamefont{Portal}},
  \bibinfo{journal}{JETP Lett.} \textbf{\bibinfo{volume}{79}},
  \bibinfo{pages}{166} (\bibinfo{year}{2004}).

\bibitem[{\citenamefont{Pogosov et~al.}(2005)\citenamefont{Pogosov, Budantsev,
  Lavrov, Plotnikov, Bakarov, and Toropov}}]{JETP.81.462}
\bibinfo{author}{\bibfnamefont{A.~G.} \bibnamefont{Pogosov}},
  \bibinfo{author}{\bibfnamefont{M.~V.} \bibnamefont{Budantsev}},
  \bibinfo{author}{\bibfnamefont{R.~A.} \bibnamefont{Lavrov}},
  \bibinfo{author}{\bibfnamefont{A.~E.} \bibnamefont{Plotnikov}},
  \bibinfo{author}{\bibfnamefont{A.~K.} \bibnamefont{Bakarov}},
  \bibnamefont{and} \bibinfo{author}{\bibnamefont{Toropov}},
  \bibinfo{journal}{JETP Lett.} \textbf{\bibinfo{volume}{81}},
  \bibinfo{pages}{462} (\bibinfo{year}{2005}).

\bibitem[{\citenamefont{Pogosov et~al.}(2002)\citenamefont{Pogosov, Budantsev,
  Uzur, Nogaret, Plotnikov, Bakarov, and Toropov}}]{PhysRevB.66.201303}
\bibinfo{author}{\bibfnamefont{A.~G.} \bibnamefont{Pogosov}},
  \bibinfo{author}{\bibfnamefont{M.~V.} \bibnamefont{Budantsev}},
  \bibinfo{author}{\bibfnamefont{D.}~\bibnamefont{Uzur}},
  \bibinfo{author}{\bibfnamefont{A.}~\bibnamefont{Nogaret}},
  \bibinfo{author}{\bibfnamefont{A.~E.} \bibnamefont{Plotnikov}},
  \bibinfo{author}{\bibfnamefont{A.~K.} \bibnamefont{Bakarov}},
  \bibnamefont{and} \bibinfo{author}{\bibfnamefont{A.~I.}
  \bibnamefont{Toropov}}, \bibinfo{journal}{Phys. Rev. B}
  \textbf{\bibinfo{volume}{66}}, \bibinfo{pages}{201303}
  (\bibinfo{year}{2002}).

\bibitem[{\citenamefont{Nihey et~al.}(1995)\citenamefont{Nihey, Hwang, and
  Nakamura}}]{PhysRevB.51.4649}
\bibinfo{author}{\bibfnamefont{F.}~\bibnamefont{Nihey}},
  \bibinfo{author}{\bibfnamefont{S.~W.} \bibnamefont{Hwang}}, \bibnamefont{and}
  \bibinfo{author}{\bibfnamefont{K.}~\bibnamefont{Nakamura}},
  \bibinfo{journal}{Phys. Rev. B} \textbf{\bibinfo{volume}{51}},
  \bibinfo{pages}{4649} (\bibinfo{year}{1995}).

\bibitem[{\citenamefont{Kataoka et~al.}(1999)\citenamefont{Kataoka, Ford,
  Faini, Mailly, Simmons, Mace, Liang, and Ritchie}}]{PhysRevLett.83.160}
\bibinfo{author}{\bibfnamefont{M.}~\bibnamefont{Kataoka}},
  \bibinfo{author}{\bibfnamefont{C.~J.~B.} \bibnamefont{Ford}},
  \bibinfo{author}{\bibfnamefont{G.}~\bibnamefont{Faini}},
  \bibinfo{author}{\bibfnamefont{D.}~\bibnamefont{Mailly}},
  \bibinfo{author}{\bibfnamefont{M.~Y.} \bibnamefont{Simmons}},
  \bibinfo{author}{\bibfnamefont{D.~R.} \bibnamefont{Mace}},
  \bibinfo{author}{\bibfnamefont{C.-T.} \bibnamefont{Liang}}, \bibnamefont{and}
  \bibinfo{author}{\bibfnamefont{D.~A.} \bibnamefont{Ritchie}},
  \bibinfo{journal}{Phys. Rev. Lett.} \textbf{\bibinfo{volume}{83}},
  \bibinfo{pages}{160} (\bibinfo{year}{1999}).

\end{thebibliography}
\end{document}